\documentclass[fleqn,10pt]{wlscirep}
\usepackage[T1]{fontenc}

\usepackage{bm}
\usepackage{rotating}
\usepackage{subfigure}
\usepackage{multirow}
\usepackage{amsmath,amssymb}

\title{
Ecology in the digital world of Wikipedia
}

\author[1,*]{Fumiko Ogushi}
\author[2,+]{J\'{a}nos Kert\'{e}sz}
\author[3,4,$\dagger$]{Kimmo Kaski}
\author[5,6,$\star$]{Takashi Shimada}

\affil[1]{Center for Mathematical Modeling and Data Science, Osaka University, Toyonaka, Osaka, 560-8531, Japan}
\affil[2]{Department of Network and Data Science, Central European University, Quellenstrasse 51, 1100 Vienna, Austria}
\affil[3]{Department of Computer Science, Aalto University School of Science, P. O. Box 15500, Espoo, Finland}
\affil[4]{The Alan Turing Institute, British Library, 96 Euston Road, London NW1 2DB, US}
\affil[5]{Mathematics and Infomatics Center, The University of Tokyo, 7-3-1 Hongo, Bunkyo-ku, Tokyo, 113-8656, Japan}
\affil[6]{Department of Systems Innovation, Graduate School of Engineering, The University of Tokyo} 

\affil[*]{ogushi@sigmath.es.osaka-u.ac.jp}
\affil[+]{kerteszj@ceu.edu}
\affil[$\dagger$]{kimmo.kaski@aalto.fi}
\affil[$\star$]{shimada@sys.t.u-tokyo.ac.jp}

\begin{abstract}
Wikipedia, a paradigmatic example of online knowledge space is organized in a collaborative, bottom-up way with voluntary contributions, yet it maintains a level of reliability comparable to that of traditional encyclopedias. The lack of selected professional writers and editors makes the judgement about quality and trustworthiness of the articles a real challenge. Here we show that a self-consistent metrics for the network defined by the edit records captures well the character of editors' activity and the articles' level of complexity. Using our metrics, one can better identify the human-labeled high-quality articles, e.g., ``featured'' ones, and differentiate them from the popular and controversial articles.
Furthermore, the dynamics of the editor-article system is also well captured by the metrics, revealing the evolutionary pathways of articles and diverse roles of editors. We demonstrate that the collective effort of the editors indeed drives to the direction of article improvement.
\end{abstract}
\begin{document}

\flushbottom
\maketitle
%
%
\thispagestyle{empty}

\section*{Introduction}
After the invention of writing and printing we are currently witnessing the third communication revolution, which is digitally-mediated and resulting in the omnipresent availability of what mankind has ever intellectually produced~\cite{Communication2015}. Human knowledge has rapidly moved to the Internet that is accessible to everyone contributing to the massive data deluge, so that our problem ``how to access information'' has turned to ``how to select information''. Any improvement in our ability to orient and/or navigate in this ever-increasing knowledge space is of great value~\cite{Dimitrov2019}.  

Wikipedia \cite{WP}, as the largest online encyclopedia, is a paradigmatic example of such a collective knowledge space that is based on ``wisdom of crowds''. The successful model of Wikipedia is that volunteers collaboratively edit the articles, improving their quality in a self-organized manner~\cite{Broughton2008,Taha2013}. The entire mechanism of Wikipedia, including the editing, the organization of editors and the categorization of the articles, is based on a bottom-up process of world-wide human creativity. Thus it is an extraordinarily appealing field of research due to its size, complex structure and -- last but not least -- because its full content and history is well documented and publicly accessible. Wikipedia is presently available in 285 languages, with 19 languages having more than one million articles.

In this paper we restrict ourselves to investigate the largest, i.e. the English language Wikipedia having currently more than 6.23 million articles (January 2021). Naturally, this huge number, together with the system of free editing result in a large variability in the length, depth and overall quality of the articles. In order to orientate the readers in this plethora of information, Wikipedia has introduced a labelling system of the articles such that some of them are literally labelled as ``good'' or ``featured'', and some are listed as ``controversial issues''\cite{ControversialArticle}, while the majority remain unlabelled.
This categorization is based on human judgement and follows the overall bottom-up principle. There has been an effort to automatize the identification of ``featured'' articles using machine learning technique~\cite{Lipka2010} and of controversial ones by means of revert statistics~\cite{Yasseri2012, Gandica2014}. These approaches are in general useful, but they tell little about the level of sophistication of the articles, which should be reflected in the complexity of the language~\cite{Lipka2010,Yasseri2012b}.  However, as regards to the content of an article, there is much more to it~\cite{Carsetti2013}. In particular, the evolution of the quality of articles deserves special attention as it should shed light onto the mechanism of how the articles improve as a consequence of the collaborative effort of editors. The understanding of this mechanism could pave the way to automated prediction of high quality articles.
 
In complex systems an adequate characterization of the constituents is often a formidable task because of the interactions, which - even if they are just binary - get extended to the entire system due to the interaction chains. For example, this explains the difference between the degree of a node in a complex network and its PageRank~\cite{Gleich2015}. The former is a single node property, while the latter takes into account the importance of the other nodes as well and can be calculated iteratively in a self-consistent manner, resulting in often a much better characterization of a node's importance. Similar reasoning in a different context has lead to the DebtRank model~\cite{Battiston2012}, which evaluates banks from the point of view of systemic risk.

Wikipedia defines a bipartite graph~\cite{Barabasi2016} with the set of the articles and the set of the editors, where a link between an editor and an article means that the former has worked on the latter. Bipartite graphs are ubiquitous in various complex contexts, examples being the networks of authors and their papers, words and sentences they occur in, or countries and goods they produce for international trade. Recently an interesting attempt has been put forward for the case of the bipartite network of the international trade to characterize the countries and their products~\cite{Hidalgo2009,Tacchella2012}. Here the concepts of ``product complexity'' and ``country fitness'' were introduced and iteratively calculated. On one hand the idea is that the fitness of a country depends on the diversity and the complexity of the products it can provide to the international market and, on the other hand, a product's complexity depends on how fit the countries have to be in order to be able to produce it. This framework has turned out to be very useful in categorising countries according to their fitness and even to make predictions about their evolution~\cite{Cristelli2013, Tacchella2018}.

In the bipartite graph of Wikipedia not only the articles have a broad distribution in quality but also the editors are very diverse in their interest, knowledge, editing skills, and devotion to Wikipedia~\cite{Yun2016}. Some of them are specialists, others work on many articles, some are highly knowledgeable and others are less so and, of course, their impact on the articles can be largely different. This is why the characterization of the articles' level of sophistication or complexity cannot be simply measured by the number of different editors working on them. Inspired by the earlier studies of a different system~\cite{Hidalgo2009,Tacchella2012,Carsetti2013} and taking into account the above mentioned inherent features of Wikipedia we define the complexity of an article and the scatteredness of an editor, and determine them in a self-consistent manner. The aim of the present study is to show that these new concepts are meaningful for the characterization of the bipartite network of Wikipedia. Moreover, we investigate how the complexity of the articles evolves in time and serve as a means to identifying high quality articles.  

\section*{Results}
\subsection*{Self-consistent metrics for the editors and articles}
Let us define a metric that enables us to rank the articles and the editors more sensitively than the local characteristics like degrees of the nodes of the network, i.e., the number of editors working on an article and the number of articles edited by a single editor or the strengths of the nodes, i.e., the total number of edits made on an article and the total number of edits carried out by an editor, respectively. On one hand, this metric should reflect the indirect effect that can be taken into account in a self-consistent way and, on the other hand, it should be automatically determinable without text analysis.
This is a non-trivial task. The level of sophistication or complexity of an article is expected to increase by the number of editors working and the number of edits made on it (i.e., the degree and the strength of a node, respectively).
However, the contribution of the editors can be very different from this point of view. Extremely active editors, who monitor a very large number of articles have less energy or time to improve the article significantly, and there are specialists who deal with articles in their expertise only and can thus contribute in a more substantial way. This bears some similarities with the problem of world trade~\cite{Tacchella2012,Carsetti2013}, but there are also marked differences.

Wikipedia can be considered as a weighted and directed bipartite graph or network that consists of $N_e$ editors and $N_a$ articles. If an editor $\epsilon$ has made an edit on article $\alpha$, there is a link $\epsilon \to \alpha$ and the number of such edits constitutes the weight of that link. 
The unweighted version of bipartite graph can be represented by the binary $N_e \times N_a$ matrix $B$:
\begin{equation}
B:\ b_{\epsilon \alpha} = 
    \begin{cases}
    1 & \mbox{(if editor $\epsilon$ has edited article $\alpha$)}\\
    0 & \mbox{(otherwise)}
    \end{cases},
\end{equation}
and the corresponding weighted version by the weighted matrix $W$:
\begin{equation}
W:\ w_{\epsilon \alpha} = \mbox{total number of edits made by editor $\epsilon$ on article $\alpha$}.
\end{equation}

As shown in Table \ref{table_netinfo} in the Methods section, the binary and weighted networks of Wikipedia are characterized by broad degree and strength distributions, respectively, as the standard deviations are larger than the corresponding averages. Both networks exhibit positive nestedness as compared to the configuration model with the same degree and strength distributions~\cite{Munoz2013}.
In what follows we will mainly use the bipartite network based on the $1,000$ most active editors but the results for the most active $2,000$ and $5,000$ editors are found to be similar. In order to increase computational efficiency we trim the network by taking into account the articles that have the degree (i.e. the number of editors on them) $10$ or more for the following analysis. For details on the selection see the Methods section.

To start we rank the articles by the number of editors $k_a$ that have contributed by editing them (see Table~\ref{table_TopDegreeArticles}). As expected, this simple degree centrality captures some features of the articles as it identifies many of the {\it prominent} ones. According to this measure, seven articles out of the top 10 turn out to be ``popular'' articles, i.e. the ones in the top 100 list of the total page views over the period from 2007 to the present (Jan. 2020)~\cite{PopularArticle2020}. However, the strong correlation between the degree centrality and popularity raises some problems. For example the popularity can result from the article being ``controversial'' or affected by news or fashion. In fact, 6 articles out of the 7 ``popular'' articles from the above list are ``controversial'', while only 2 are labelled ``featured'' and 3 ``good''.
Thus we can conclude that the high degree rank is more characteristic for the article's ``controversiality'' and ``popularity'' than for its quality or complexity. This problem could not be resolved by using instead of the degree, the strength rank calculated from the weighted matrix $W$. 
In addition, we have tried to incorporate the indirect effect of nodes at a distance by using the ranking based on the eigenvector centrality~\cite{Newman2003} of the projection of the bipartite network onto the set of articles. 
However, this does not improve the situation considerably, as shown in Table~\ref{table_TopDegreeArticles} in the Methods section.

The impact of the editors on the goodness of an article is expected to be a non-monotonic function of the activity and the diversity of articles an editor deals with. Very low activity editors, who edit one or a few articles a few times only have usually little impact on the goodness of an article. On the other hand, very active editors monitoring and contributing to a large number of articles (e.g., admins) are again less responsible for the contextual improvements as their role is more maintenance.
As we consider the top $1,000$ editors, all of them turn out to edit thousands of articles. Some of the edits are substantial additions to complex matters and some can be more maintenance type edits such as small or systematic corrections. Without text analysis it is hard to identify the maintenance like activities. However, we can expect that an editor working on a very large number of articles (some editors, indeed, edit millions), especially including articles with low ``goodness'' (or complexity) values, is less likely to make substantial content edits to improve the complexity of articles,
as their capacities (time and energy) are scattered among too many items.
On the other hand, active editors working on relatively few articles, high quality or complexity are expected to add essential contributions to the articles by their edits.
Therefore, we define the {\it scatteredness} $D_i$ of an editor $i$, as the harmonic sum of the article complexities he or she edits. The complexity of an article is then naturally defined as a harmonic sum of the scatteredness values of the editors who edited the article, which is essentially the sum of the editors' contributions to the contents writing.

In order to differentiate the goodness or complexity of an article from its popularity and controversiality and to have an adequate measure to characterise also the editors, we introduce the following self-consistent metrics: 
\begin{equation}
    \tilde{D}_i^{(n+1)}
    = \sum_{\alpha} \frac{\displaystyle w_{i\alpha}}{\displaystyle C_\alpha^{(n)}},
	\quad
	\tilde{\mbox{$C$}}_j^{(n+1)}
	= \sum_{\epsilon} \frac{\displaystyle w_{\epsilon j}}{\displaystyle D_\epsilon^{(n)}},
	\qquad
	\left(
	\mbox{for binary network,}
	\
	\tilde{\mbox{$D$}}_i^{(n+1)}
    	= \sum_\alpha \frac{\displaystyle b_{i\alpha}}{\displaystyle C_\alpha^{(n)}},
	\quad
	\tilde{\mbox{$C$}}_j^{(n+1)}
	= \sum_{\epsilon} \frac{\displaystyle b_{\epsilon j}}{\displaystyle D_\epsilon^{(n)}}
	\right)
	\label{eq_def_fitness_1}
\end{equation}
\begin{equation}	
	D^{(n+1)}_i =
	\frac{\tilde{D}^{(n+1)}_i}{\displaystyle \frac{1}{N_e} \left( \sum_\epsilon \tilde{D}^{(n+1)}_\epsilon \right)},
	\quad
	C^{(n+1)}_j =
	\frac{\displaystyle \tilde {C}^{(n+1)}_j}{\displaystyle \frac{1}{N_a} \left( \sum_\alpha \tilde{C}^{(n+1)}_\alpha \right)}.
	\label{eq_def_fitness_2}
\end{equation}
Here $D_i^{(n)}$ is the scatteredness of the editor $i$, and $C_j^{(n)}$ is the complexity of the article $j$, as calculated after $n$ times of recursive evaluation steps. Note that after each re-evaluation step, Eqs. (\ref{eq_def_fitness_1}) we normalize the scatteredness and complexity measures to get Eqs. (\ref{eq_def_fitness_2}).

\begin{figure}[htbp]
\center
\includegraphics[width=1.0\linewidth]{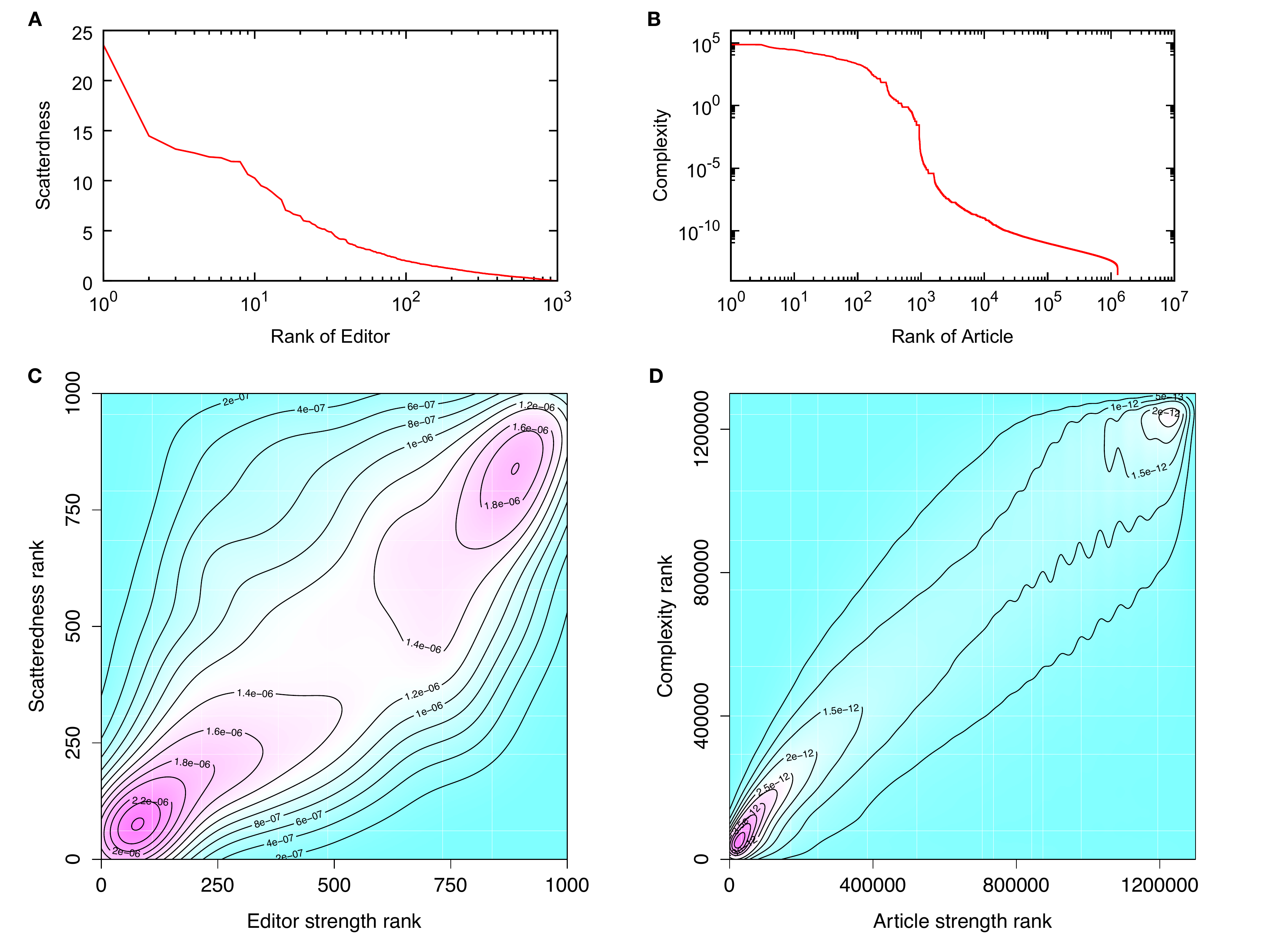}
\caption{The distribution of the scatteredness of editors and the complexity of articles calculated from the self-consistent measures in Eqs (\ref{eq_def_fitness_1}, \ref{eq_def_fitness_2}).
({\bf A}) The rank plot of the scatteredness of editors,
({\bf B}) the rank plot of the complexity of articles,
({\bf C}) the density distribution of editors in the rank-rank plot of the strength $s_\epsilon = \sum_{\alpha} w_{\epsilon \alpha}$ and the scatteredness, and
({\bf D}) the density distribution of articles in the rank-rank plot of the 
strength $s_\alpha = \sum_{\epsilon} w_{\epsilon \alpha}$ and the complexity.
}
\label{fig:fitness_complexity} 
\end{figure}

Starting from the uniform initial condition $D^{(0)}_i = 1, \ C^{(0)}_j = 1$, the proposed recursive process yields good convergence for both the scatteredness and complexity. In addition, the obtained distributions turn out to be smooth, as depicted in Fig.~\ref{fig:fitness_complexity}. Although the scatteredness of editors and the complexity of articles are correlated with the corresponding degrees, they show considerable variations.
Because we adopt a harmonic sum in our metric, one can expect that the result is sensitive to the threshold we set for the article degree to be taken into account for the analysis.
However, we confirm that the good convergence and the smoothness of the score distributions are kept for different threshold values, and the performance of the score is the best for the chosen threshold value as discussed in detail in the Methods Section and in the SI.

\subsection*{Characterization of articles by the self-consistent complexity measure}
\begin{figure}[tbh]
\includegraphics[width=1.0\linewidth]{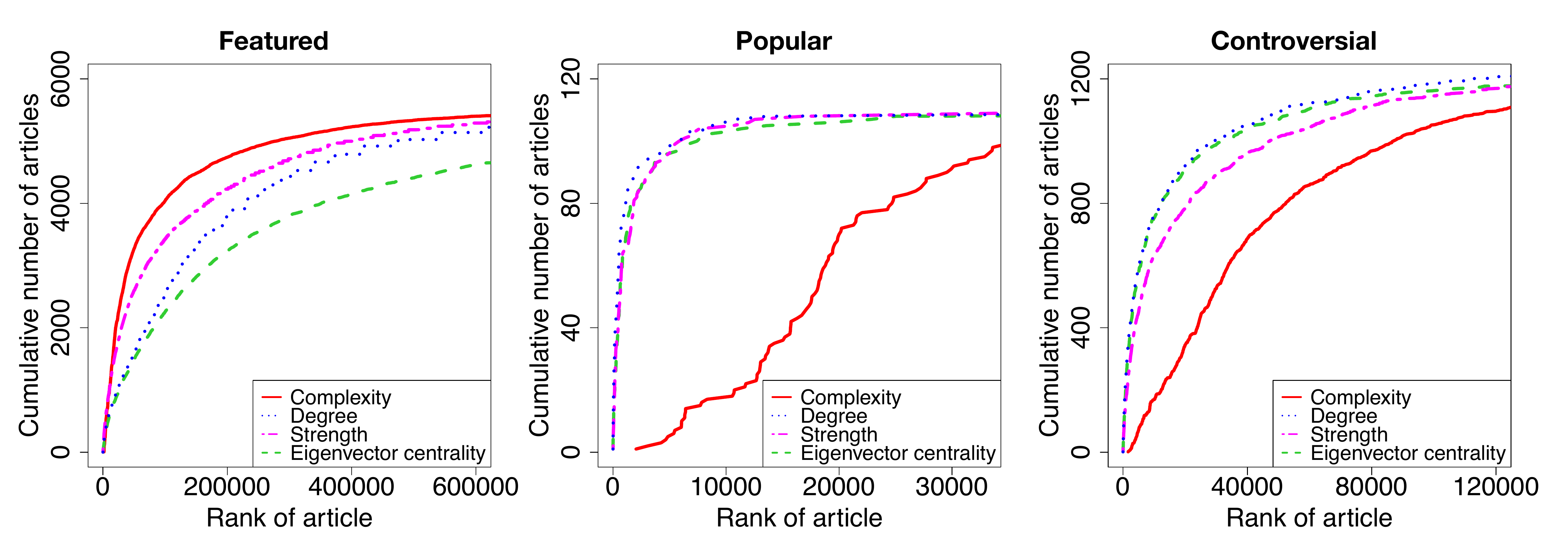}
\caption{Cumulative number of ``featured'', ``popular'', and ``controversial'' articles contained in the top-ranked or top-N hit articles for Complexity, Degree, Strength, and Eigenvector centrality measures. The Complexity works best for finding the ``featured'' articles, while it picks the ``popular'' articles and the ``controversial'' articles more selectively, than the other measures.
}
\label{fig:cumN_fa_top1k} 
\end{figure}

\begin{figure}[htbp]
\includegraphics[width=1.0\linewidth]{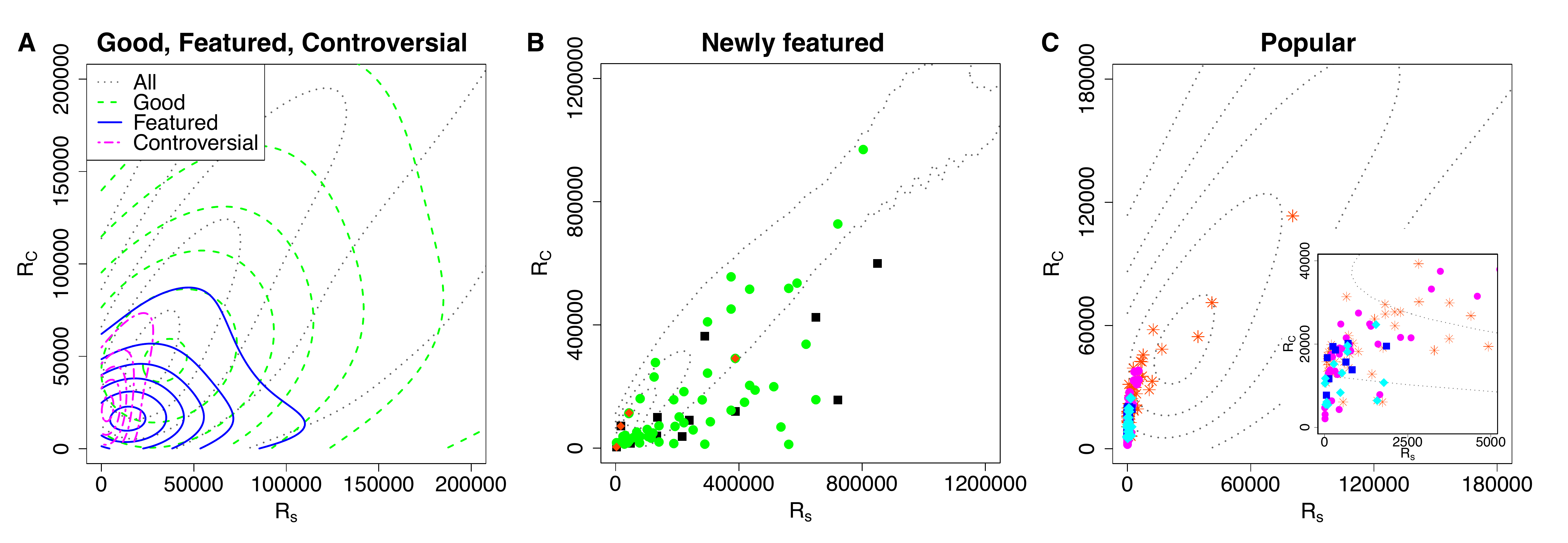}
\caption{
Distributions of labeled Wikipedia articles in the strength rank vs. complexity rank plane.
({\bf A}) Contour plot of the density distributions of
all (gray dotted line), ``good'' (green dashed line), ``featured'' (blue solid line), and ``controversial'' (magenta dot-dashed line)
articles. In comparison to the density distribution of all as well as ``good'' articles, the ``featured'' and ``controversial'' articles are accumulated in the region of higher ranks, i.e. where both $R_s$ and $R_C$ are small, while the ``good'' articles have lower ranks and both $R_s$ and $R_C$ larger. The ``featured'' articles tend to have higher complexity rank than all articles for its strength, while ``controversial'' articles tend to have extremely high strength rank but smaller complexity rank than ``featured'' articles. 
({\bf B}) In the $R_s$ vs. $R_C$ plane the positions of ``newly featured'' articles that were labelled ``featured'' within the period from 20th September 2019 to 20th February 2020. 
Green circles and black squares are the ``newly featured'' articles promoted from the ``good'' articles and non-labeled articles, respectively, shown on top the contour (gray dotted) lines of the density distribution of all articles. The ``newly featured'' articles are mainly promoted from the ``good'' articles, the distribution of which is concentrated in the lower triangle of the $R_s$ vs. $R_C$ plane compared to the distribution of all articles.
The original positions of ``newly featured'' articles are found further to the right, compared to the overall distribution of ``good'' articles, which indicates a strong tendency that articles with higher complexity rank relative to strength rank correlate with the higher chance for them to obtain ``featured'' label. This tendency is stronger for the originally non-labeled articles.
The ``newly featured'' articles that are also labelled as ``emerging popular'' articles (red diamonds), having earned top $5,000$ page views around the observation period, tend to be located in the upper triangle region, i.e. the strength rank is higher than the complexity rank.
({\bf C}) In the $R_s$ vs. $R_C$ plane the positions of articles labelled ``popular'' (orange star), ``popular'' \& ``featured'' (blue squares), ``popular'' \& ``controversial'' (magenta circles), and ``popular'' \& ``featured'' \& ``controversial'' (cyan diamonds), shown on top of the contour (gray dotted) lines of the density distribution of all articles. These ``popular'' singly, doubly and triply labelled articles are localized in the region of high strength rank and relatively high complexity rank, such that ``popular'' \& ``controversial'' labelled articles correlate with higher strength rank while ``popular'' \& ``featured'' labelled articles correlated with the higher complexity rank.
}
\label{fig:density_of_labeled_articles}
\end{figure}

Let us investigate how the complexity measure characterizes the articles labelled ``featured'', ``good'', ``popular'', and ``controversial'' in comparison with other measures. In Fig.~\ref{fig:cumN_fa_top1k} we compare the efficiency of the different measures by plotting the cumulative number of identified articles with different labels as a function of the article's rank as calculated from the measure. It is seen in the left panel that the complexity measure $C$ works better for finding the ``featured'' articles than the other commonly used measures, i.e., the degree, the strength, and the eigenvector centrality. From this figure it is also evident that the complexity measure turns out to pick up selectively the ``popular'' articles (middle panel) and the ``controversial'' articles (right panel). 
This distinguishing feature of the complexity measure is also seen in the different distributions of the labeled articles of the strength rank -- complexity rank plane, as depicted in Fig.~\ref{fig:density_of_labeled_articles} A.
In this plot we observe that in comparison to all as well as ``good'' articles, the ``featured'', and ``controversial'' articles share a common characteristic by both of them being concentrated in the top-rank area (bottom-left corner) of the strength rank -- complexity rank plane. Moreover, the ``controversial'' articles appear most strongly concentrated, with strength rank considerably higher than the complexity rank, while the ``featured'' articles are distributed more evenly but concentrated slightly to higher complexity ranks.

Next we turn our attention to the differences between the complexity rank and the strength rank by focusing on the ``newly featured'' articles, i.e. the articles that were labeled as “featured” within the period from 20th September 2019 to 20th February 2020, having been before that ``non-labelled'' or ``good'' articles.
As depicted in Fig.~\ref{fig:density_of_labeled_articles} B, the ``newly featured'' articles that emerged mainly from ``good'' but some also from ``non-labelled'' articles have a clear tendency for the complexity rank being higher than the strength rank. On the other hand the ``newly featured'' articles that were around the above-mentioned time period also ``emerging popular'' articles (i.e. the ones that earned top 5,000 page views in 2019),
tend to have the strength rank higher than the complexity rank.
In Fig.~\ref{fig:density_of_labeled_articles} C we show the distribution of articles labelled ``popular'', ``popular'' \& ``featured'', ``popular'' \& ``controversial'', and ``popular'' \& ``featured'' \& ``controversial'' in the strength rank-- complexity rank plane. All of these articles are located in the upper triangular part of the plane, thus implying that their strength rank is clearly higher than their complexity rank. However, some of the doubly and the triply labelled articles show relatively high complexity rank. This observation supports the view that the higher rank in the complexity metric $C$ serves as an indicator of the character of an article. On the other hand the high strength (but also degree) rank of the ``featured'' articles is partly due to a ``pre-featured'' or ``post-featured'' process that might make the finding of ``featured'' articles by the complexity measure $C$, difficult. Some possible examples of such processes are further proof-reading edits of the article before nomination (pre-process) or the article becoming ``popular'' or ``controversial'' because of the ``featured'' label (post-process).

\begin{figure}[htbp]
\begin{center}
\includegraphics[width=0.49\linewidth]{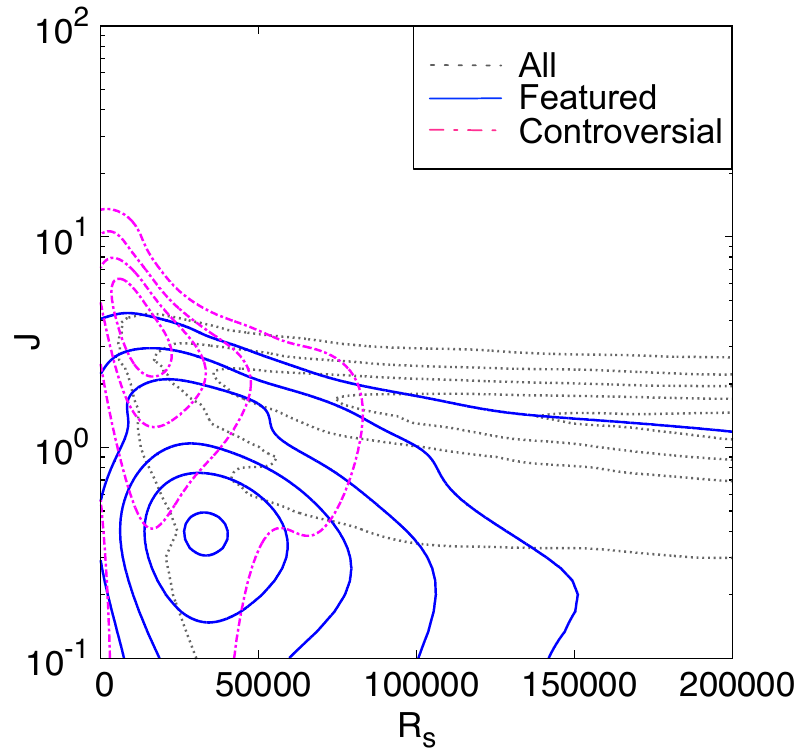}
\caption{
Distribution of the ``featured'', and ``controversial'' articles in the $R_s$ (strength rank)-$J$ (complexity-strength rank ratio) plane. The ``featured'' articles are distributed in the small $J < 1$ region and well separated from the ``controversial'' articles which are peaking at $J > 1$ region.
}
\label{fig:density_j}
\end{center}
\end{figure}

As discussed above and shown in Fig.~\ref{fig:density_of_labeled_articles} A and C, the ``featured'' and ``controversial'' articles share a common characteristic, as their distributions in the strength rank -- complexity rank plane partly overlap but are still distinguishable.
In order to better differentiate their distributions we introduce the following rank ratio $\displaystyle J_\alpha = \frac{R_{C_\alpha}}{R_{s_\alpha}},$
where $R_{C_\alpha}$ and $R_{s_\alpha}$ are the ranks of the article $\alpha$ according to the complexity measure and to the strength in the descending order, respectively. This complexity - strength rank ratio basically evaluates the average scatteredness of the editors of an article, relative to other articles with similar strength. In Fig.~\ref{fig:density_j} we plot the complexity - strength rank ratio vs. the strength rank, and indeed the distribution of the ``featured'' articles become quite well separated from the distribution of the ``controversial'' articles. Furthermore, we observe that for the ``featured'' articles typically the complexity - strength rank ratio gets values $J<1$, which indicates that complexity rank values exceed those of strength rank, while for the ``controversial'' articles this is overturned and the ratio gets typically higher values $J>1$, as expected.

\subsection*{Dynamics of articles in Wikipedia}
\begin{figure}[htbp]
\begin{center}
\includegraphics[width=18cm]{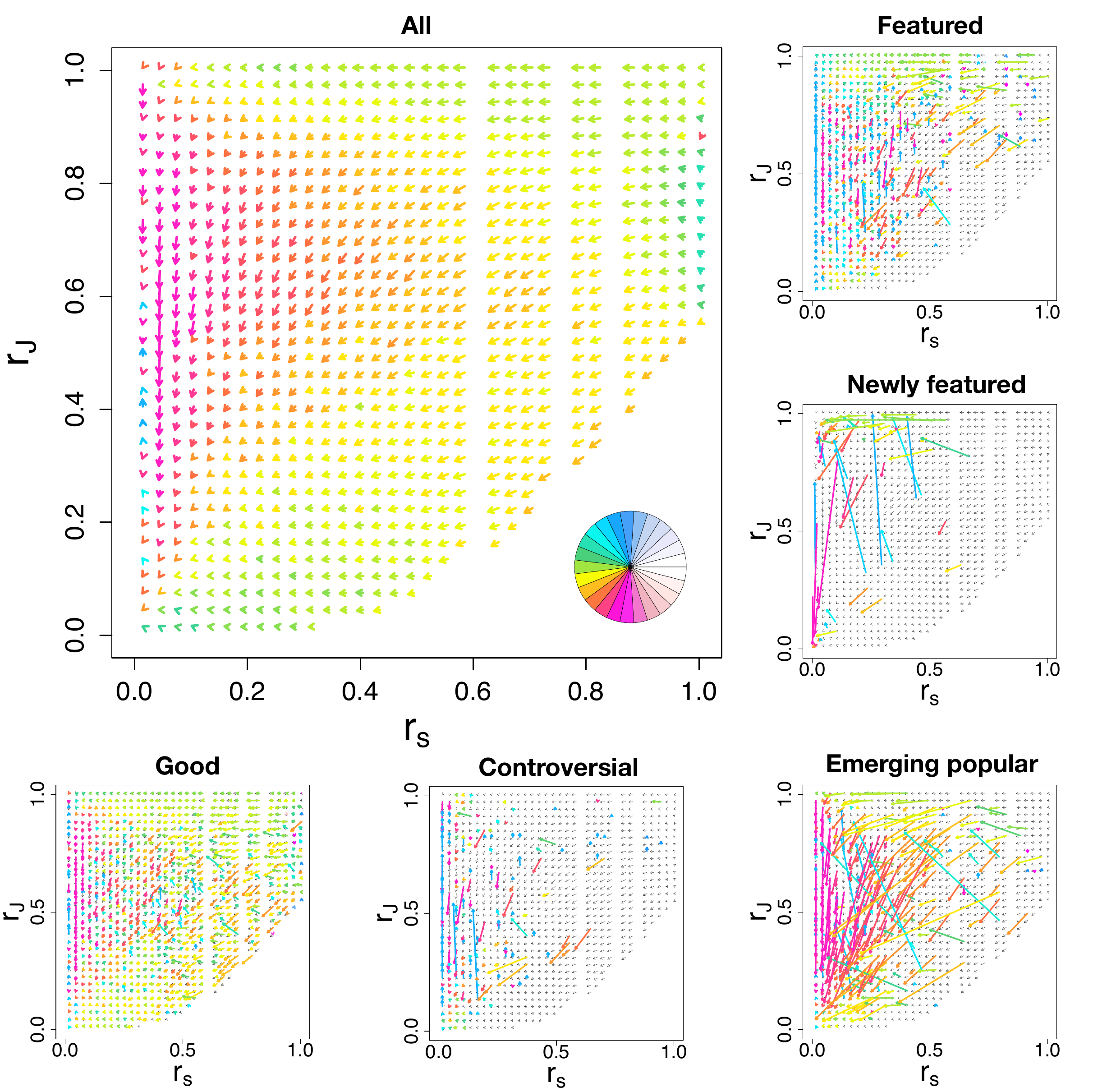}
\caption{
The temporal evolution of articles in the plane of relative ranks of the strength $r_s = R_s/N_a$, and of the complexity - strength rank ratio $r_J = R_J/N_a$, for the period ranging from 20th Sep 2019 to 20th Feb 2020. 
({\bf Top left}) Average flows of all articles, i.e. both labelled and non-labelled articles. 
({\bf Top right to bottom right and then horizontally to the left}) Flows of articles labeled as ``Featured'', ``Newly featured'', ``Emerging popular'', ``Controversial'', and ``Good''.
Arrows are colored according to their directions, as shown in the top left panel.
Each arrow for the ``Newly featured'' articles corresponds to each of the $70$ articles without averaging. 
For other panels, the average flows are plotted. The flows of labeled articles are overlaid on the average flow of all articles (grey).
Note that in the low strength regime, many articles have the same strength and hence have the tie rank, which appears as the vertical blank strips in the flow plot.
}
\label{fig_rankchange}
\end{center}
\end{figure}

In the previous subsection we demonstrated the relevance of our complexity measure in characterizing and ranking the articles of different categories or labels. Here we focus on the temporal evolution of the articles in terms of their relative ranks of the strength, $r_s = R_s/N_a$, and of the complexity - strength rank ratio $J$, $r_J = R_J/N_a$. 
As shown in the main panel of Fig. \ref{fig_rankchange}, the Wikipedia articles evolve in time basically by gaining strength.
Also it turns out that an article getting new edits would lead to an increase in its complexity, unless the scatterdnesses of some of the already connected editors shoot up during the observation time. The rate of the complexity increment of an article is determined by the average of the inverse scatterdness ($D_\epsilon^{-1}$) of the editors who recently edited it.
The relation between this rate and the original strength is well visualized in the average flow pattern (see the main panel in Fig. \ref{fig_rankchange}). Here we see that there is an overall leftward stream with down trend, i.e. the rank in the complexity does not get higher when compared to the change of rank in strength. This implies that, as an article gets more edits, it becomes increasingly difficult to keep getting even more edits from the less scattered editors (i.e. specialists).
In the panel of relative ranks of all articles in Fig. \ref{fig_rankchange} it is seen that the overall evolutionary flow of labelled and non-labelled articles does not go only towards a single ``sink'' in the bottom left corner with both the relative strength rank $r_s$ and complexity - strength rank ratio $r_J$ small, but towards a broad area in the left boundary with $r_s$ small but $r_J$ variable. This reflects the fact that the average gain of complexity per edit tends to relax to a mature value with considerable diversity.

In order to see the role of articles with different labels (i.e. ``featured'', ``newly featured'', ``good'', ``controversial'', and ``emerging popular'') in the overall evolutionary flow of articles we present them as separate small panels in Fig. \ref{fig_rankchange}. Here it is evident that the flow patterns of labelled articles exhibit clear differences in reflection of their evolutionary path from non-labelled stage to different labelled stages. 
Here we also see that the ``featured'' and ``newly featured'' articles have low-$J$ rank, especially in the low-strength region. However, the
``featured'' articles in the high-strength region show an upstream trend, meaning that their acquiring contents-edit contributions tend to be kept, even in their matured stage.
The movements of ``newly featured'' articles is found to be stochastic in nature, due to not being averaged. 
However, these movements tend to keep $J$ low and some of them show a very drastic upward motion i.e. a rapid growth in complexity. As for the articles labelled ``Good'' we observe their evolution to be quite similar to the evolution of all articles, but showing a slightly enhanced upward trend (due to contents edits), especially in the high-strength (matured) region. In the case of the ``controversial'' articles we observe them mostly concentrated in the high-strength-rank and high-J-rank region as already shown in Fig. \ref{fig:density_j}. In this region the pattern of flow bears some similarity with the flow patterns of ``featured'' and ``good'' articles as well as to some extent with all articles. Outside this region the flow pattern looks rather chaotic. 
Of all the labelled articles the ``popular'' articles are found to have most stable flow pattern over the period of observation and to be concentrated in the bottom-left triangle region in the $r_s$ vs. $r_J$ plane (not shown). 
In contrast the evolution of ``Emerging popular'' articles, constructed from the top $5,000$ page views in 2019, show rapid growth in the strength and the downward trending flow pattern to the bottom-left corner the $r_s$ vs. $r_J$ plane. 
This behavior might be explained as the high popularity for a shorter time span was achieved because it became controversial and hence needed maintenance edits, or, inversely, it became controversial in terms of the high need of maintenance because of its intrinsic emerging popularity.

As is mentioned above, some ``controversial'' articles show eye-catching upward move like the ``featured'' articles, while some others move down like ordinary articles or even drastically like the ``emerging popular'' articles. This might reflect the fact that there are different reasons for articles to be controversial. If the controversy is mainly from the impact or delicateness of that topic to the public, the editing activity of ``reverting war'' and the maintenance effort should be similar to that of the majority ``controversial'' articles and ``emerging popular'' articles. If the controversy is of the multifaceted nature or the difficulty in settling correct explanation of that issue, updates to fix the article need much expertise of that field and hence the activity becomes more like the one of ``featured'' articles.
In fact, five out of top-10 upward moving articles in the top 1\% strength area in the $r_s-r_J$ plot, namely, {\it sexually transmitted infection, disability, eugenics, tobacco,} and {\it string theory}, are about the category of ``Science, Biology, and Health'', which occupies only about 10\% of ``controversial'' articles. Moreover, seven out of top 10 ``controversial'' articles in the increase of complexity 
(namely, {\it Chiropractic, Aspartame, Aspartame controversy, Alternative medicine, Homeopathy, Vaccine hesitancy}, and {\it Ebola virus disease})
are from ``Science, Biology, and Health'', while only two are from the category of ``People'' that constitutes about 30\% of ``controversial'' articles.

Also in the ``newly featured'' articles, we see some counter flows (downward motion). Most conspicuous ones are found to be {\it Horologium} and {\it Portrait of Mariana of Austria}. It is due to their rapid growth in strength with slight increase in complexity. In the edit history of these articles one can find that there were intense editing by active editors just before the nomination of those articles for ``featured'' articles, which is a good example of above argued possible ``pre-featured'' process.
The reason why these particular articles needed such a process seems to be that those are in the disciplines which require special fine formats to be a qualified article (constellation and paintings, respectively). These investigations on the counter flows in labeled articles show the usefulness of our complexity measure and the derivative index of complexity - strength rank ratio $J$ to characterize the articles, even beyond the article labeling.

\section*{Discussion}
We have shown that in analysing the quality of the Wikipedia articles and their evolutionary paths the complexity measure $C$ and the derivative measure of complexity - strength rank ratio $J$ serve as sensitive indicators.
The performance of $C$ and $J$ for finding ``featured'' articles with less picking ``popular'' and ``controversial'' articles is better than other measures which also use only the information of editing network, i.e. the degree, strength, and eigenvector centrality.
However, the main significance of our study is in discovering more holistically the ecology of Wikipedia, i.e., how the different types of editorial activities influence the complexity of the articles and its evolutionary dynamics. 
The usefulness of the complexity measure $C$ in finding ``featured'' articles implies that the main contribution to the complexity of an article is not due to the most active editors, whose contributions are scattered among many articles but rather due to those non-top editors, who focus in their edits on contributing on substance. 
This means that, although the editing activity is heavily dominated by the top-editors, their role is mainly devoted to maintain the Wikipedia system, while the improvement of the complexity of an article is expected from editors with more specialized interest. 

A novel aspect of our work is the systematic study of the evolution of articles in the complexity - strength rank space in Fig.~\ref{fig_rankchange}. 
The basic assumption of Wikipedia that the articles continuously improve by the collective effort of the editors has been challenged repeatedly~\cite{TedPappas2004}.
The overall trend of the articles towards the high complexity - high strength corner in Fig.~\ref{fig_rankchange} is a quantitative evidence for this basic assumption of Wikipedia.
The apparent differences in the flow diagrams with regard to the different characters or labeling of the articles is a further indication of the sensitivity of our approach to separate articles with high editorial activity according to their level of sophistication. 

Our self-consistent formulation for analysing the quality of Wikipedia articles share features with the analysis of countries' economic fitness and the complexity of products they provide for the world trade network \cite{Tacchella2012, Cristelli2013}. However, there are fundamental differences between these two eco-systems, which stem from differences in the underlying processes and dependence on available resources (for comparison in terms of the metrics used, see the SI). While the economic ecosystem is dependent on countries' limited natural and human resources, the Wikipedia ecosystem depends on rather unlimited collective knowledge space of human creativity by individual editors. Their individual choices or selectiveness in editing articles gives insight into their behavioural and activity patterns. As discussed above, the selective edit records, focusing mainly on {\it good} articles, can be regarded as a good indication of some editors' tendency to make contents writing contribution to the article, which in our analysis framework is (inversely) measured by the {\it scatterdness} of individual editors. 
Finally, as we expect that this kind of bipartite mechanisms might be quite universal in other human collaborative, creative or social networks, they would be interesting further targets for our self-consistent analysis approach.

\section*{Methods}
\subsection*{Data Preparation}
The data we use in this study consists of the edit records on individual articles, which were dumped from {\it Wikimedia Downloads} \cite{Wikimedia} over a period ranging from the 20th Sept. 2019 till the 20th Feb 2020.
We exclude bot or bot-like editors and non-content articles from our analysis, by removing bots listed in Wikipedia bot \cite{Bot}, unregistered anonymous users, and editors who edit mainly non-article pages.
Articles labeled as ``redirect'' or ``disumbiguation'' are also excluded from our analysis.
The basic characteristics of the Wikipedia edit network after this pre-processing are shown in TABLE~\ref{table_netinfo}.
\begin{table}[htbp]
  \caption{Characteristics of the editor-article unweighted ($B$) and weighted ($W$) networks for the top $1,000$ (the actual number of editors $N_e$ after removing the bot-like ones is $984$), $2,000$ ($N_e = 1,974$), and $5,000$ ($N_e = 4,951$) editors.
  For the top $1,000$ editors, the characteristics of the network after trimming the low-degree articles ($k_A < 10$) is shown together.
  $\langle k_E \rangle$ and $\sigma_{k_E}$ are the average and the standard deviation of the editors' degree, and $\tilde{\eta}^E_k$ is the average nestedness of the editors' links relative to a configuration model with the same parameters ($\langle k_E \rangle$ and $\sigma_{k_E}$, see SI).
  The corresponding quantities for the links of articles are denoted by $\langle k_A \rangle, \sigma_{k_A}$, and $\tilde{\eta}^A_k$, respectively.
  For the weighted networks ($W$), we show the average ($\langle s_E \rangle, \langle s_A \rangle$) and the standard deviation ($\sigma_{s_E}, \sigma_{s_A}$) of the strength, as well as the relative nestedness quantities calculated using the link weights,   $\tilde{\eta}^E_s$ and $\tilde{\eta}^A_s$.
  }
  \begin{tabular}{rr|rrrrrr|rrrrrr}
    Data & 
    & & & $B$ & & &
    & & & $W$ & & &
    \\
  	\ $N_{e}$ & \quad $N_{a}$&
  	$\langle k_E \rangle$ & $\sigma_{k_E}$ & $\tilde{\eta}_{k}^E$
  	&
	$\langle k_A \rangle$ & $\sigma_{k_A}$ & $\tilde{\eta}_{k}^A$
    &
    $\langle s_E \rangle$ & $\sigma_{s_E}$ & $\tilde{\eta}_{s}^E$ &
	$\langle s_A \rangle$ & $\sigma_{s_A}$ & $\tilde{\eta}_{s}^A$
	\\
	\hline
   	$984$ & $5{,}417{,}280$ &
	 $4.30$ & $7.92$ & $1.52$ & $7.80$ & $9.77$ & $1.34$
	 &
	 $9.37$ & $12.3$ & $1.01$ & $17.0$ & $43.9$ & $2.00$
	 \\
   	(trimmed) & $1{,}272{,}018$ &
	 $2.41$ & $3.79$ & $1.15$ & $18.6$ & $15.3$ & $1.19$ &
	 $5.81$ & $6.33$ & $0.751$ & $45.0$ & $81.7$ & $1.60$
    \\
	& &
	& & & & & &
	& & & & &
	\\
   	$1{,}974$ & $5{,}481{,}212$ &
	$2.65$ & $5.85$ & $1.67$ & $9.53$ & $13.4$ & $1.46$
	&
	$6.19$ & $9.31$ & $1.17$ &
	$22.3$ & $63.0$ & $2.35$
	\\
   	$4{,}951$ & $5{,}552{,}014$ &
	$1.38$ & $3.85$ & $1.86$ &
	$12.3$ & $20.3$ & $1.69$
	&
	$3.40$ & $6.33$ & $1.42$ &
	$30.3$ & $91.1$ & $2.85$
	\\
	& &
	{\small $(\times 10^4)$} & {\small $(\times 10^4)$} & & & & &
	{\small $(\times 10^4)$} & {\small $(\times 10^4)$} & & & &
	\\
  \end{tabular}
  \label{table_netinfo}
\end{table}

In order to calculate the self-consistent metric for the scatteredness of editors and the complexity of articles, we construct a network composed of articles that have been edited by $10$ or more editors. 
Articles with extremely low complexity have typically very few editors and are generated systematically and then left almost intact. 
Examples of such would be every settlement in a certain state, and every (sub) species in a certain taxon, and lists of those.
However, filtering articles by their ``list'' tags does not work well because some articles with list tags contain rich information and some other articles essentially contain only list information but without list tags. 

As shown in TABLE~\ref{table_netinfo},
the only remarkable change in the network characteristics after the trimming is that the nestedness of the weighted network in the editor side becomes smaller than that of the configuration model.
This implies a division-of-labour aspect of the editing activity on articles edited by many editors (ten or more, here). 
Also the top $10$ articles in $degree$ rank are listed with the article titles, labels, and rank in the Eigenvector centrality measure, as presented in  TABLE~\ref{table_TopDegreeArticles}. While all of those are prominent articles, their prominence are more associated with the popularity and controversiality of the article.

\begin{table}[htbp]
  \caption{Top $10$ articles in degree rank. Most of the top articles are ``controversial''
  and ``popular'', i.e. listed in the top $100$ most popular articles for the period from 2007 to 2020.}
  \label{table_TopDegreeArticles}
  \centering
  \begin{tabular}{c|l|c|c|c|c|cc}
        Degree Rank: $r_{k_a}$ &\multicolumn{1}{c|}{Article title} & Featured & Good & Controversial & Popular & (Eigenvector Centrality Rank) 
       \\ 
	\hline 
	$1$ & Barack Obama & $\checkmark$ & $-$& $\checkmark$ & $\checkmark$ & 1
	\\
	$2$ & Donald Trump & $-$ & $-$ & $\checkmark$ & $\checkmark$ & 2
	\\
   	$3$ & Wikipedia & $-$ & $-$ & $\checkmark$ & $\checkmark$ & 28
	\\
	$4$ & United States &  $-$ & $\checkmark$ & $\checkmark$ & $\checkmark$ & 12
	\\
	$5$ & George W. Bush & $-$ & $\checkmark$ & $\checkmark$ & $-$ & 10
   	\\
	$6$ & Adolf Hitler & $-$ & $\checkmark$ & $\checkmark$ & $\checkmark$ & 37
	\\
	$7$ & Michael Jackson & $\checkmark$ & $-$ & $\checkmark$ & $\checkmark$ & 20
	\\
	$8$ & Chicago & $-$ & $-$ & $\checkmark$ & $-$ & 6
	\\
	$9$ & London & $-$ & $\checkmark$ & $-$ & $\checkmark$ & 19
	\\
	$10$ & Paris & $-$ & $\checkmark$ & $-$ & $-$ & 5
	\\
	\hline 
  \end{tabular}
\end{table}

\subsection*{Calculating self consistent measures}
In order to obtain the scatteredness of editors and the complexity of articles, we iterate the calculation Eq. (\ref{eq_def_fitness_1}) and (\ref{eq_def_fitness_2}) until it converges around a fixed point. 
We stop the iteration when the following convergence condition is fulfilled:
\begin{equation}
\delta D^{(n)}
=
\sqrt{
\frac{1}{N_e}
\sum_{\epsilon=1}^{N_e} \left( D^{(n)}_\epsilon - D^{(n-1)}_\epsilon \right)^2
}
\
< 10^{-4}.
\end{equation}
A typical number of iteration steps we need for the convergence is less than $100$.
The scores obtained from the fixed point are different for the non-trimmed network from that of the trimmed network, while the good convergence and the smoothness of the score distributions remain. However, the change in the resulting score ranking is rather discontinuous against the change in the threshold value and the performance of the score for the non-trimmed network is worse in the later discussed sense. Therefore, our choice of the threshold value at $10$ is not arbitrary (see SI for detail).

\bibliography{ref}

\section*{Acknowledgements}
FO thanks for hospitality of Aalto University.
JK and KK acknowledge support from EU HORIZON 2020 INFRAIA-2019-1 (SoBigData++) No. 871042. KK also acknowledges the Visiting Fellowship at The Alan Turing Institute, UK.
TS was partly supported by JSPS KAKENHI grant number 18K03449.

\section*{Author contributions statement}
All authors (FO, JK, KK, and TS) conceived the study plan. FO and TS conducted the data preparation and the numerical calculation. All authors analysed the results and participated in writing the paper.

\section*{Additional information}
\textbf{Competing interests} 
The authors have no competing interests.


\clearpage
\section*{Supplemental Information}

\subsection*{Network characteristics}
\subsubsection*{Degree and Strength}
We denote the degree of editor $\epsilon$ and article $\alpha$ as
\begin{equation}
    k^E_\epsilon = \sum_\alpha b_{\alpha \epsilon}, \quad k^A_\alpha = \sum_\epsilon b_{\alpha \epsilon},
\end{equation}
and the strength of those as
\begin{equation}
    s^E_\epsilon = \sum_\alpha w_{\alpha \epsilon}, \quad s^A_\alpha = \sum_\epsilon w_{\alpha \epsilon}.
\end{equation}

\subsubsection*{Measuring Nestedness}
In order to characterize the nestedness of our Wikipedia edit-network, we need a measure that is independent of degree distribution. For this reason, we adopt a nestedness measure, which was proposed for ecological networks \cite{Munoz2013}.
For our network $a_{\epsilon \alpha} = B_{\epsilon \alpha}$ for the binary case and $a_{\epsilon \alpha} = W_{\epsilon \alpha}$ for the weighted case, the (local) nestedness of the edits made by editor $i$ and that made by editor $j$ is defined as follows
\begin{equation}
    \eta^E_{ij} = \frac{\displaystyle \sum_{\alpha = 1}^{N_A} a_{i\alpha} a_{j\alpha}}{k^E_i k^E_j},
\end{equation}
which basically counts the overlap between these two edit patterns.
In the same manner, the nestedness between the edit patterns on article $i$ and $j$ reads
\begin{equation}
	\eta^A_{ij} = \frac{\displaystyle \sum_{\epsilon = 1}^{N_E} a_{\epsilon i} a_{\epsilon j}}{k^A_i k^A_j}.
\end{equation}
Note that the degree of editors and articles ($k^E_i, k^A_j$) in the equations above should be substituted by the corresponding strength ($s^E_i, s^A_j$) for the weighted network $W$.
Then the local nestedness of an editor and an article are defined as the averages for those nodes as
\begin{equation}
\eta^E_i = \frac{\displaystyle \sum_{j \neq i}^{N_E} \eta^E_{ij}}{N_E-1},
\qquad
\eta^A_i = \frac{\displaystyle \sum_{j \neq i}^{N_A} \eta^A_{ij}}{N_A-1}.
\end{equation}
Then the global nestedness of the network is characterized by the averages of the editors' nestedness and the articles' nestedness:
\begin{equation}
\eta^E = \frac{\displaystyle \sum_{i, j \neq i}^{N_E} \eta^E_{ij}}{(N_E-1)N_E},
	\qquad
\eta^A = \frac{\displaystyle \sum_{i, j \neq i}^{N_A} \eta^A_{ij}}{(N_A-1)N_A}.
\end{equation}
A merit on taking this measure is that the baseline nestedness, which is defined as the nestedness of the network with no correlation among the links, keeping the degree distribution (configuration model)
\begin{equation}
a_{ij}
= \frac{k^E_i k^A_j}{\langle k_E \rangle N_E}
= \frac{k^E_i k^A_j}{\langle k_A \rangle N_A},
\end{equation}
can be simply (analitycally) calculated as
\begin{equation}
\bar{\eta}^{E_C} \equiv \langle \eta^{E_C}_i \rangle
= \left( \frac{1}{N_A} \right) \frac{\langle k_A^2 \rangle}{\langle k_A \rangle^2},
\qquad
\bar{\eta^{A_C}} \equiv \langle \eta^{A_C}_j \rangle
\left( \frac{1}{N_E} \right) \frac{\langle k_E^2 \rangle}{\langle k_E \rangle^2}.
\end{equation}
The global nestednes relative to this baseline,
\begin{equation}
\tilde{\eta}^E = \frac{\eta^E}{\bar{\eta}^{E_C}},
\qquad
\tilde{\eta}^A = \frac{\eta^A}{\bar{\eta}^{A_C}},
\end{equation}
are found to be larger than $1$ in the Wikipedia network (Table \ref{table_netinfo}). This tells that it is positively nested (i.e. the edit pattern is positively correlated).

\clearpage

\subsection*{The effect of trimming the low-degree articles}
The number of editors working on the article (i.e. article degree) influences the goodness of the article. Although the contribution of the editors on the articles can be largely different, the low-degree articles tend to contain poor information. We thus set the trimming threshold for the article degree. The network with threshold $x$ is the network after trimming the articles with degree $k_a \leq x$. The network with threshold $0$ is the original network. 

\begin{figure}[htbp]
\includegraphics[width=0.33\linewidth]{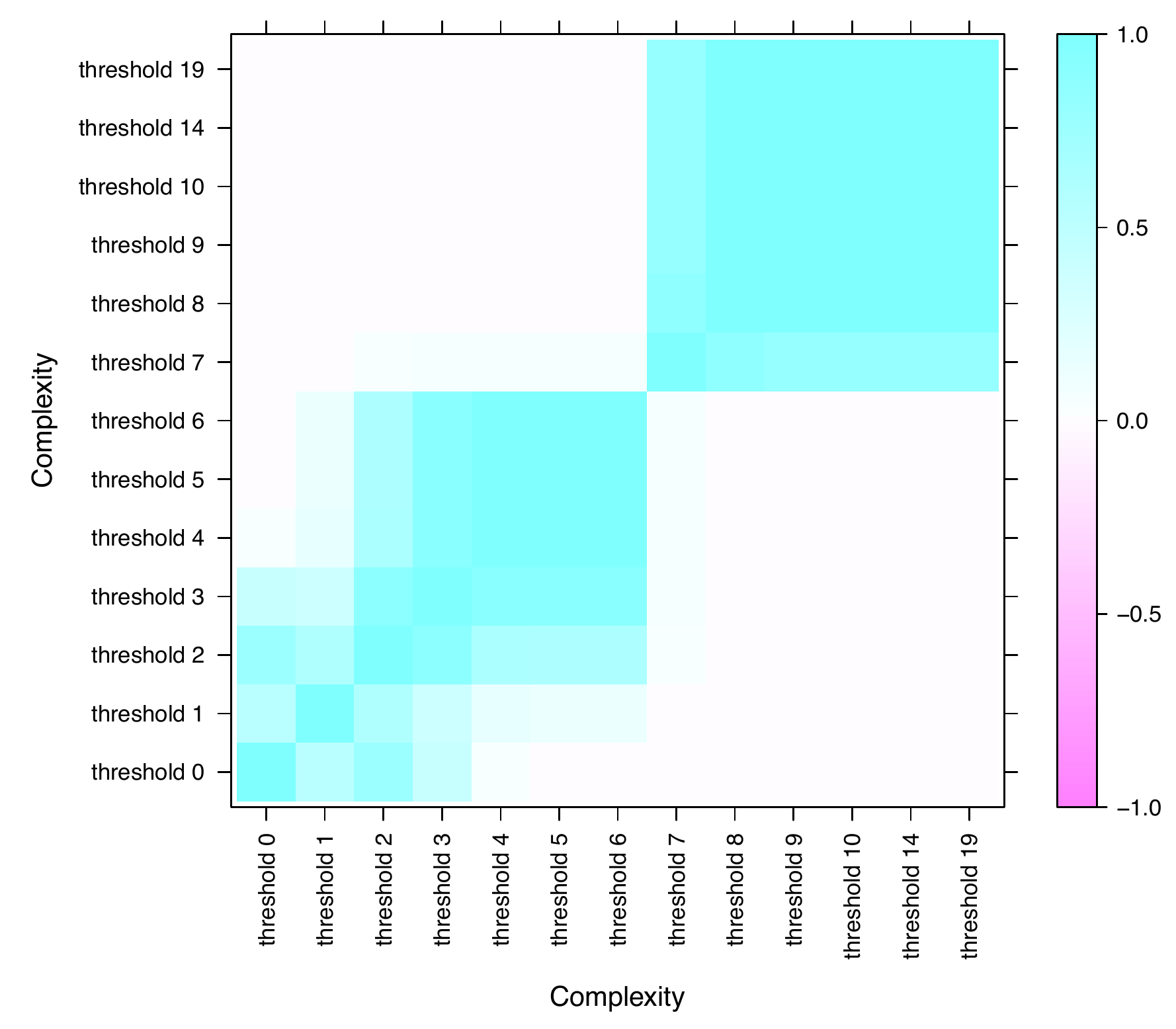}
\includegraphics[width=0.33\linewidth]{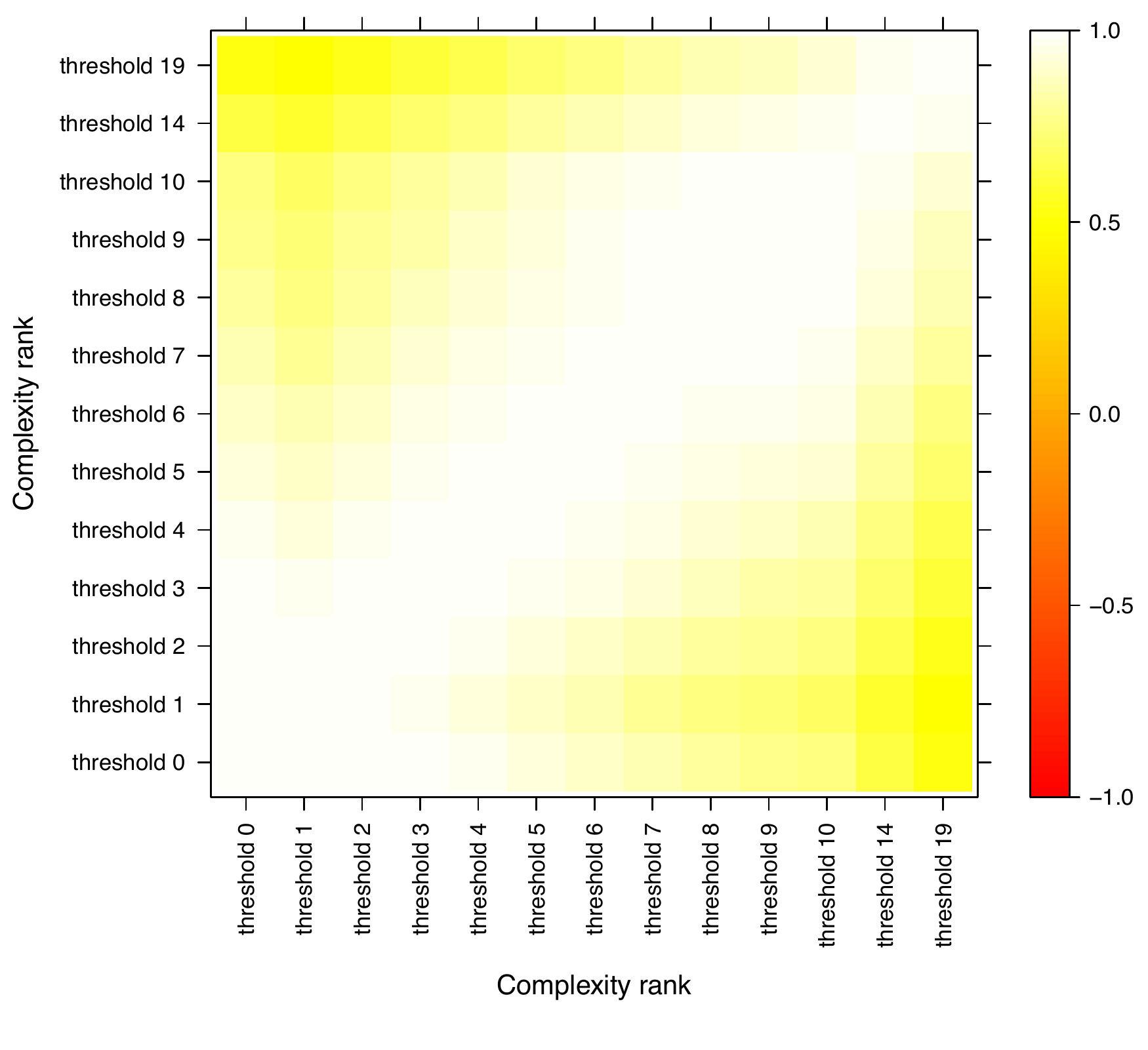}
\includegraphics[width=0.33\linewidth]{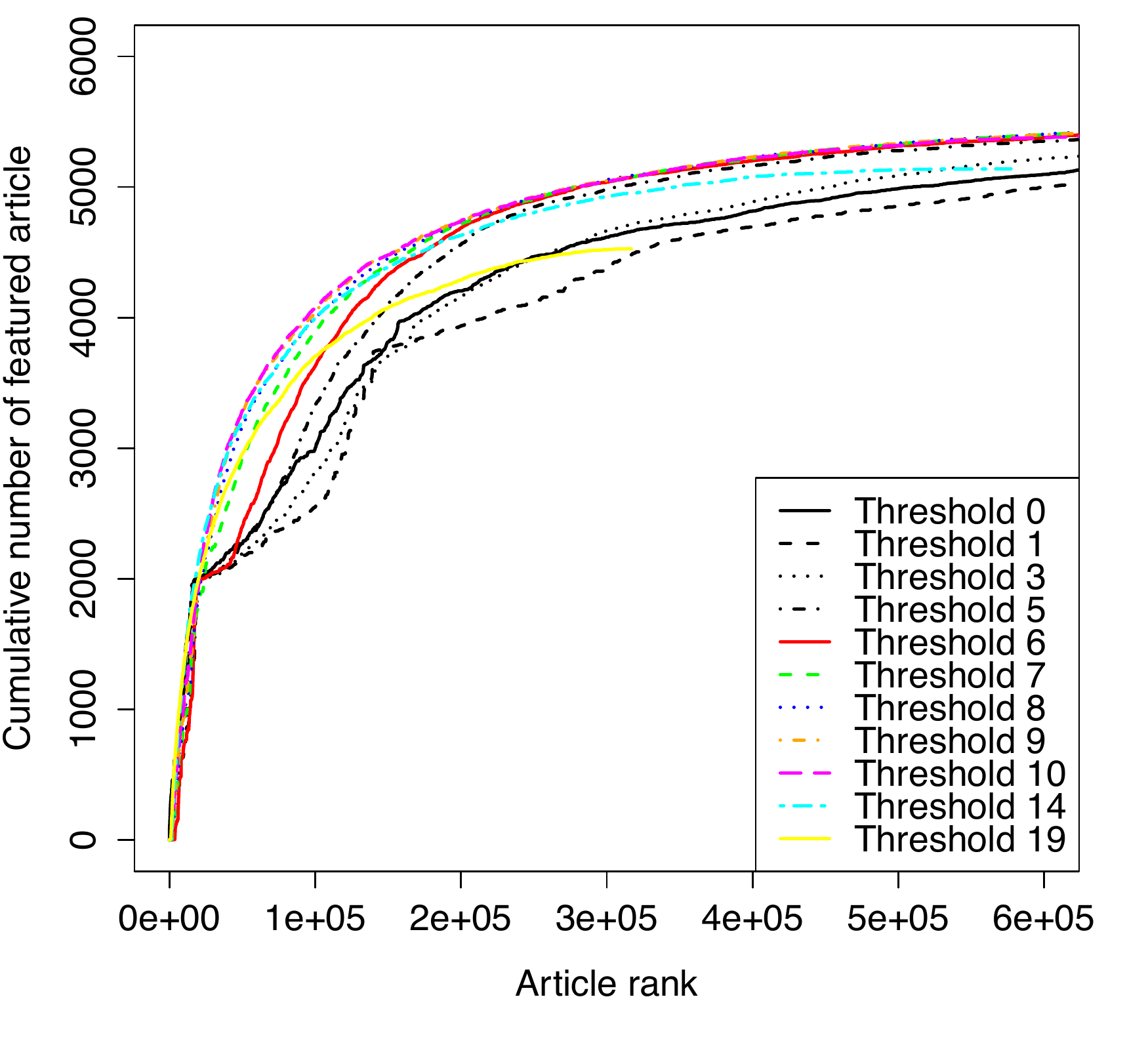}
\caption{The effect of the trimming threshold. 
(Left) Person's correlation coefficient of the complexity measure. 
(Middle) Spearman's rank correlation of the complexity rank. 
(Right) Cumulative number of ``featured'' articles contained in the top-ranked articles in complexity measure for different threshold values. To find the ``featured'' articles, the networks after trimming the articles with degree $k_a \le 9$ (orange) and $k_a \le 10$ (magenta) work best.
}
\label{SI_fig:trimming_effect}
\end{figure}

To assess the effect of the trimming threshold on our self-consistent analysis, we calculate the correlation of the complexity measure between the networks with different threshold values. 
The Pearson's correlation coefficient $p_{xy}$ of the complexity measure between two networks with threshold value $x$ and $y$ is defined as, 
\begin{equation}
 p_{xy} 
  = \frac{\sum_{\alpha}^{all}(C_{\alpha}^{x}-\langle C^{x} \rangle)(C_{\alpha}^{y}-\langle C^{y} \rangle)}
 {\sqrt{\sum_{\alpha}^{all}(C_{\alpha}^{x}-\langle C^{x} \rangle)^2}\sqrt{\sum_{\alpha}^{all}(C_{\alpha}^{y}-\langle C^{y} \rangle)^2}},
  \label{SI_eqs_Pearson}   
\end{equation}
where $C_{\alpha}^{x}$ and $C_{\alpha}^{y}$ denote the complexity of the article $\alpha$ of the network with threshold $x$ and $y$, respectively. The average complexity of the network is, by definition, $\langle C^{x} \rangle = \langle C^{y} \rangle =1$. 
The Spearman's rank correlation coefficient $\rho_{xy}$ is given by the Pearson's correlation coefficient calculated on the ranks of the data instead of the original values, 
\begin{equation}
 \rho_{x,y} 
  = \frac{\sum_{\alpha}^{all}(R_{C_{\alpha}^{x}}-\langle R_{C^{x}} \rangle)(R_{C_{\alpha}^{y}}-\langle R_{C^{y}} \rangle)}
 {\sqrt{\sum_{\alpha}^{all}(R_{C_{\alpha}^{x}}-\langle R_{C^{x}} \rangle)^2}\sqrt{\sum_{\alpha}^{all}(R_{C_{\alpha}^{y}}-\langle R_{C^{y}} \rangle)^2}},
  \label{SI_eqs_Spearman}
\end{equation}
where $R_{C_{\alpha}^{x}}$ and $R_{C_{\alpha}^{y}}$ denote the complexity rank of the article $\alpha$ of the network with threshold $x$ and $y$, respectively. $\langle R_{C^{x}} \rangle$ and $R_{C^{y}}$ denote the average rank. As shown in Fig.~\ref{SI_fig:trimming_effect}, there are three different fixed points, the networks with threshold $0 \sim 2$, the networks with threshold $3 \sim 6$, and the networks with threshold $\geq 7$. 

Top-N hit rate for finding the ``fetured'' articles depends on the threshold value. The finding accuracy does not monotonically increase with increasing the threshold value as shown in the right panel in Fig.~\ref{SI_fig:trimming_effect}. The original network skims ``featured'' articles faster than the network with threshold $1$. As increasing the threshold value from $1$ to $19$, the accuracy increases for the networks with small threshold but decreases for the networks with large threshold. The network with threshold $9$ and $10$ work best. We use the optimal threshold $9$ for our self-consistent analysis. 

\clearpage

\subsection*{Relation to the metric of country fitness and product complexity}
Our formulation is inspired by the self-consistent measure proposed for countries' economic fitness $F$ and product's complexity $Q$ \cite{Tacchella2012, Cristelli2013}
for the world trade network, calculated as
\begin{equation}
\tilde{F}_i^{(n+1)} = \sum_{\alpha} b_{i \alpha} Q_\alpha^{(n)},
\quad
\tilde{Q}_j^{(n+1)} = \left( \sum_{\epsilon}\frac{b_{\epsilon j}}{F_\epsilon^{(n)}}\right)^{-1}.
\label{SI_eqs_FitnessComplexity}
\end{equation}
The reason why we can not simply adopt the original definition of quality for the article complexity is the following.
At a first glance, one could be inclined to regard the editors as the countries and the articles as the products, naturally from its causality relation: editors write the articles and not vice versa. However, on the Wikipedia network, there is no {\it capability} which limits the touch of editors to an articles (e.g. everyone could make an edit on an article of quantum physics and mid-century history of a certain local village of Japan). Therefore the selectiveness of the editors editing an article is not a good measure of its complexity or goodness, meaning that this way of straight forward application is not appropriate for Wikipedia network.

On the contrary, the selectiveness of the opponent articles in the Wikipedia bipartite network gives good information about the editors. Because we here take the top-editors, all of them are editing on thousands of articles. Some of the edits are contents edit and some can be more maintenance type edits such as small or systematic correction. Although it is hard to distinguish these maintenance like activity from the edit size or other information, we can expect that an editor editing so many articles (some indeed edit millions), especially including articles with low ``goodness'' (complexity), has a lower probability to make a contents edit contributing the ``goodness'' of articles. The selective edit records mainly on ``good'' articles, on the other hand, can be regarded as a good indication of the higher contents writing contribution to the article by his each edit. This type of character of editors, not the ``fitness'' of it, is measured by the {\it scatterdness} in our framework.

\begin{figure}[htbp]
\includegraphics[width=0.33\linewidth]{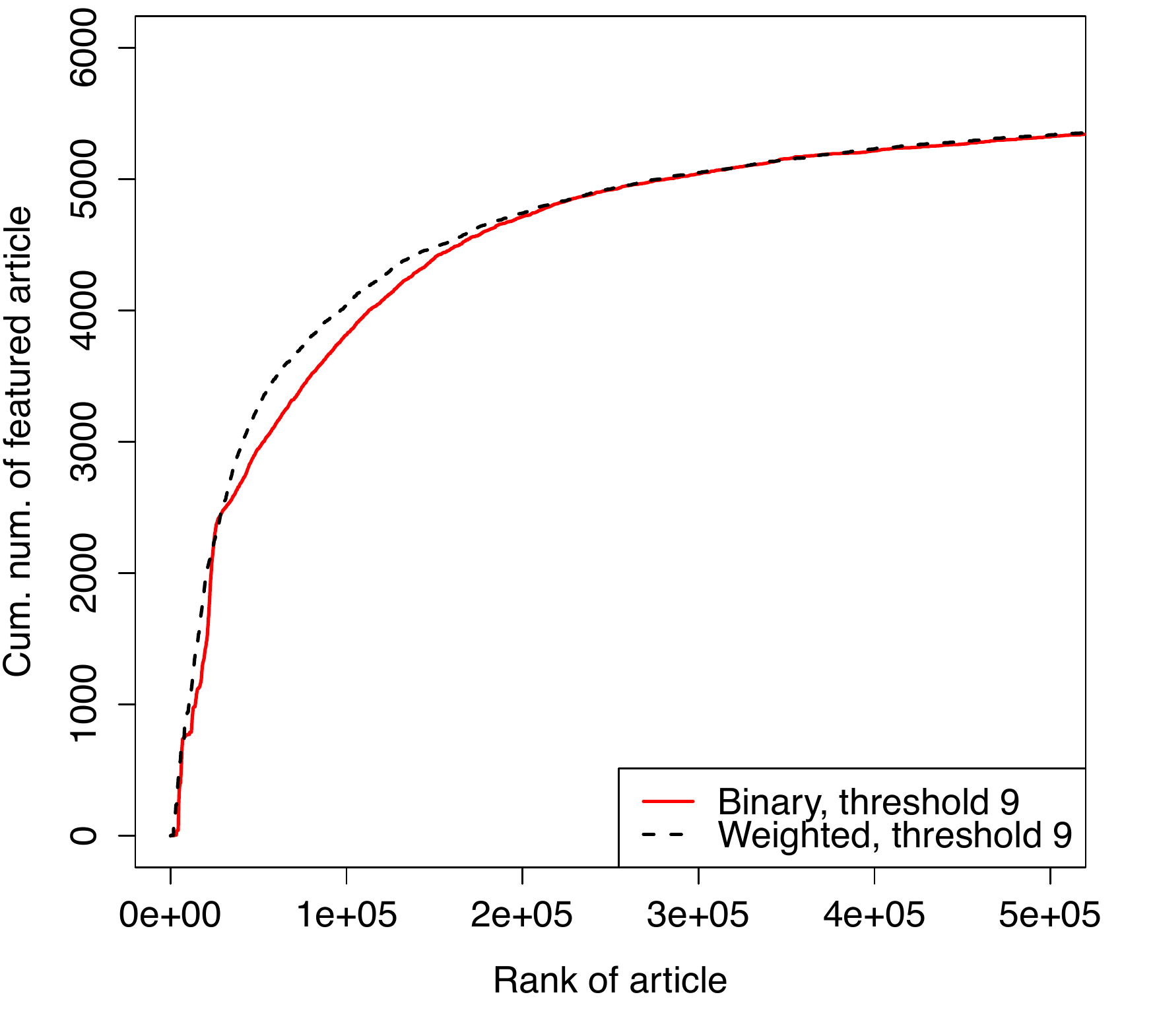}
\includegraphics[width=0.33\linewidth]{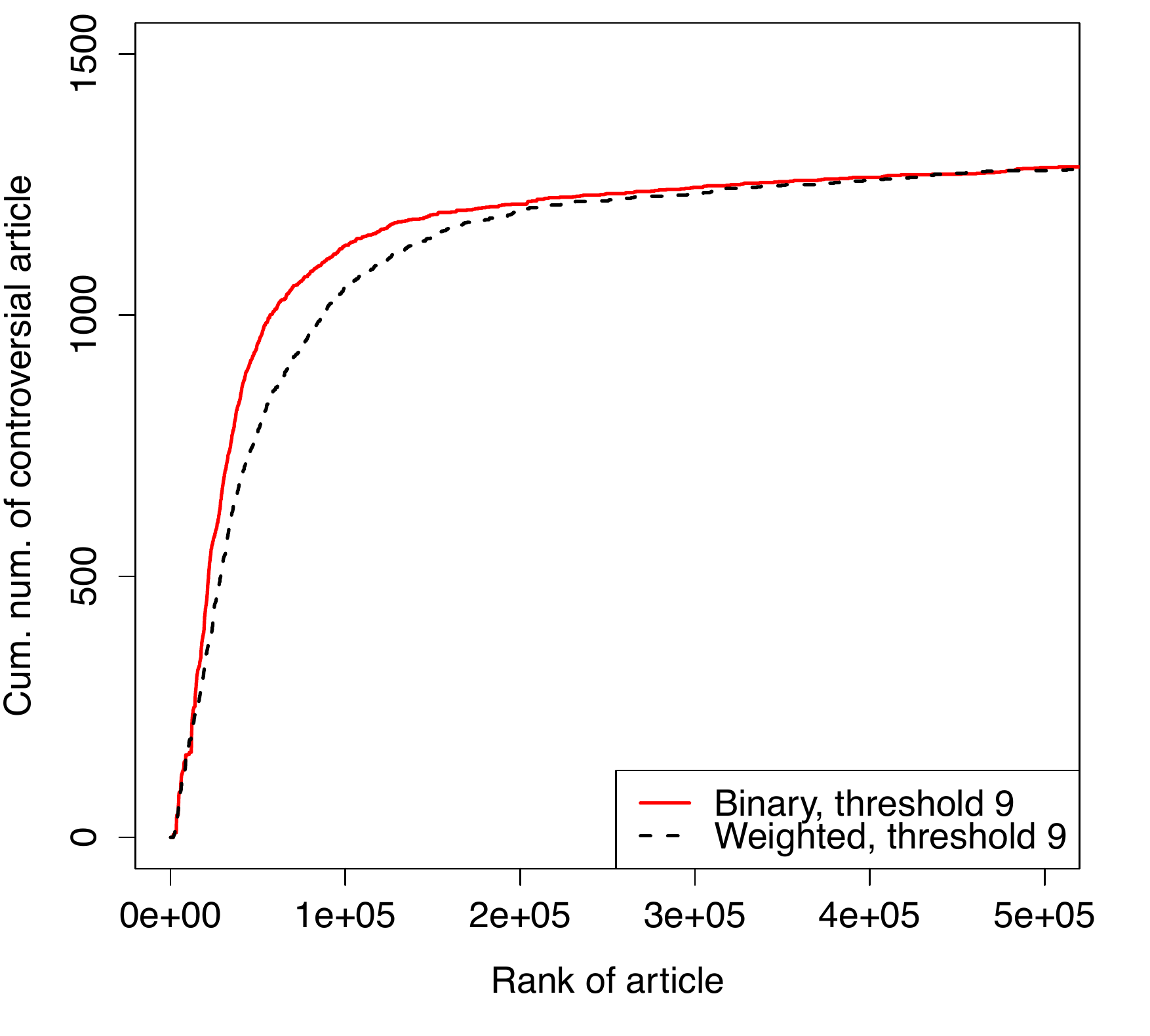}
\includegraphics[width=0.33\linewidth]{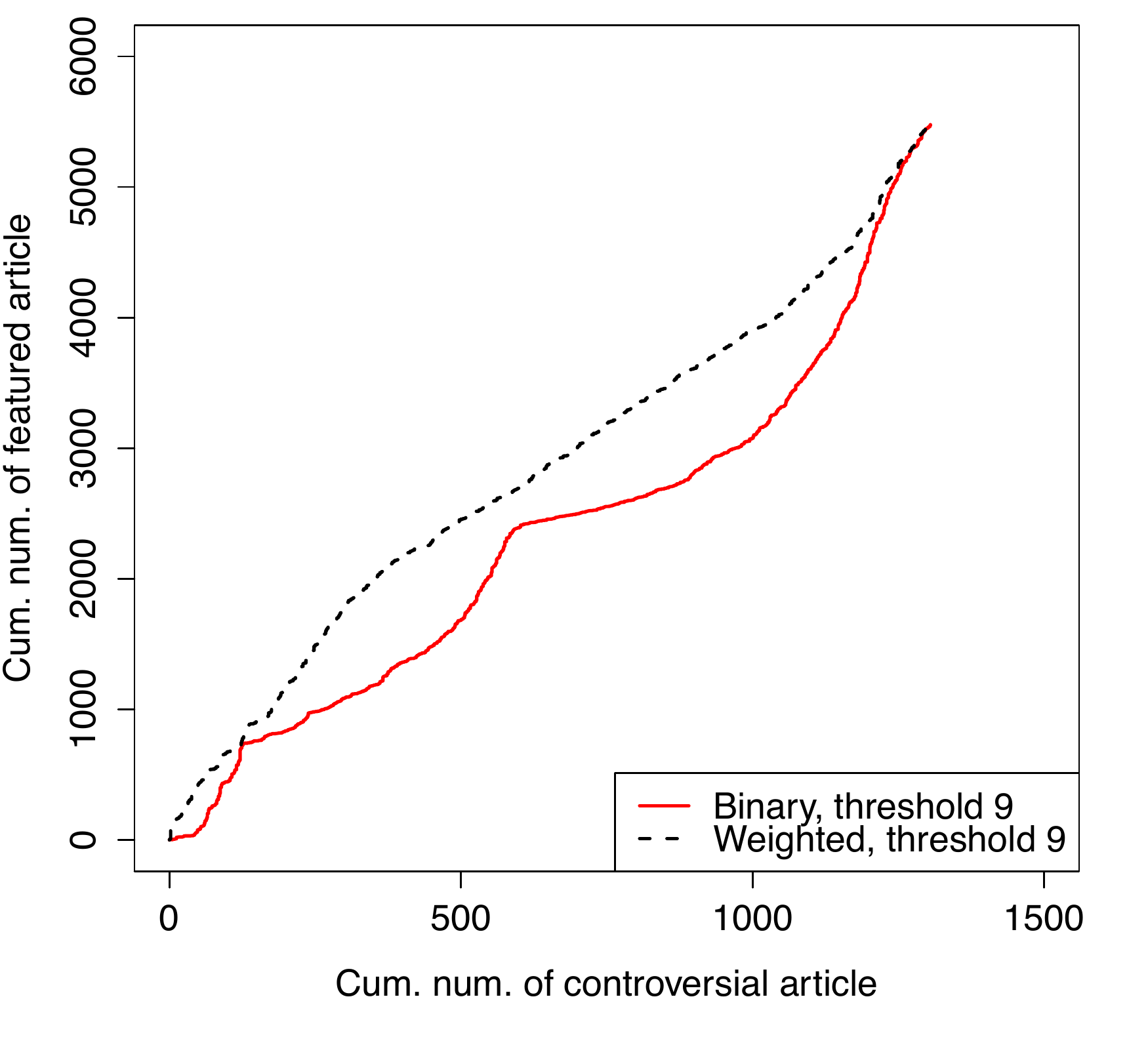}
\caption{Performance of complexity measure for the binary network $B$ and the weighted network $W$.
(Left) Cumulative number of ``featured'' articles contained in the top-ranked articles in each complexity measure.
(Middle) Cumulative number of ``controversial'' articles contained in the top-ranked articles in each complexity measure. 
(Right) The relation between the cumulative numbers of ``featured'' and ``controversial'' articles, which shows the performance of complexity measure to find ``featured'' articles without picking ``controversial'' articles.
}
\label{SI_fig:cumN_binary_weight}
\end{figure}

Note that for binary network $b_{\epsilon \alpha}$, our definition is equivalent to taking the inverse of the products' complexity index as new complexity index for articles in the original definition for economy relation (Eqs. (\ref{SI_eqs_FitnessComplexity})), i.e. $C_j = Q_j^{-1}$, and therefore essentially the same as taking the inverse rank for the complexity in the original definition, except for the normalization condition. In the present work, we take weighted network $w_{\epsilon \alpha}$ mainly because of its better performance as shown in Fig. \ref{SI_fig:cumN_binary_weight} and hence this direct relation is lost.
However, the mapping relation for the binary network still gives a good guide for considering the convergence of our non-linear recursion process to a non-trivial fixed point with smooth distribution of the resulting values, thanks to the intensive work on the convergence condition for the fitness-complexity measure\cite{Pugliese2016}.

\vspace{1cm}

\end{document}